\documentclass[aps,prl,reprint,superscriptaddress,preprintnumbers,a4paper]{revtex4-1}
\usepackage{graphicx}
\usepackage{amsmath}
\usepackage{amsfonts}
\usepackage{amssymb}

\usepackage{bm}
\usepackage{hyperref}

\def\sss{\scriptscriptstyle}

\newcommand{\Tint}[1]{{\hbox{$\sum$}\!\!\!\!\!\!\!\int\,}_{\!\!\!\!\raise-0.9ex\hbox{$\scriptstyle{#1}$}}}

\def\siml{{\ \lower-1.2pt\vbox{\hbox{\rlap{$<$}\lower6pt\vbox{\hbox{$\sim$}}}}\ }}
\def\simg{{\ \lower-1.2pt\vbox{\hbox{\rlap{$>$}\lower6pt\vbox{\hbox{$\sim$}}}}\ }}

\def\naBla{{\bm \nabla}}

\def \als {\alpha_{\mathrm{s}}}

\def \m2   {\mu^{2 \epsilon}}

\newcommand{\MS}{{\overline{\rm MS}}}
\def\siml{{\ \lower-1.2pt\vbox{\hbox{\rlap{$<$}\lower6pt\vbox{\hbox{$\sim$}}}}\ }}
\def\simg{{\ \lower-1.2pt\vbox{\hbox{\rlap{$>$}\lower6pt\vbox{\hbox{$\sim$}}}}\ }}

\def\p{{\bm p}}

\def\v{{\bm v}}
\def\x{{\bm x}}

\def \OO {\mathcal{O}}

\def\ij{{i \cdots j}}
\def\lra{\leftrightarrow}
\def\onetwo{{1\lra2}}
\def\twotwo{{2\lra2}}

\def\C{{\cal C}}

\def\S{{\cal S}}

\def\pressure{{\cal P}}

\def\j{{\bm j}}

\def\degen{\nu}
\def\dpslash{\frac{d^3\p}{(2\pi)^3} \>}

\def\qhat {\hat{q}}

\def\md {m_{\sss D}}

\def\Eq#1{Eq.~\eqref{#1}}

\def\taupi{\tau_\pi}
\def\tauj{\tau_j}
\def\kmu{k_{\mu_\alpha}}

\begin{document}
\title{
Second-order Hydrodynamics in Next-to-Leading-Order QCD
}

\author{Jacopo Ghiglieri}

\affiliation{Theoretical Physics Department, CERN, CH-1211 Geneva, Switzerland}

\author{Guy D. Moore}
\affiliation{Institut f\"ur Kernphysik, Technische Universit\"at Darmstadt\\
Schlossgartenstra\ss e 2, D-64289 Darmstadt, Germany}
\author{Derek Teaney}
\affiliation{Department of Physics and Astronomy, Stony Brook University,\\
Stony Brook, New York 11794-3800, United States}

\preprint{CERN-TH-2018-079}

\date{\today}

\begin{abstract}
  We compute the hydrodynamic relaxation times $\taupi$ and $\tauj$
  for hot QCD at next-to-leading order in the coupling
  with  kinetic theory.
  We show that certain dimensionless ratios of second-order to first-order
  transport coefficients obey bounds which apply whenever a kinetic
  theory description is possible; the computed values lie somewhat
  above these bounds. Strongly coupled theories with holographic duals
  strongly violate these bounds, highlighting their distance from a 
  quasiparticle description.
\end{abstract}

\maketitle

\textbf{Introduction:}  
The quark-gluon plasma (QGP) produced at RHIC
\cite{Adler:2003kt,Adams:2004bi} and the LHC
\cite{ALICE:2011ab,Chatrchyan:2013nka,Aad:2014fla,ALICE:2016kpq}
appears to be an excellent fluid.  Despite the small system
size, viscous hydrodynamics does a good job describing many collective
properties, spectra, and correlations
\cite{Heinz:2013th,Gale:2013da}.
To be causal and stable
\cite{Israel:1979wp,Hiscock:1985zz},
such treatments must
work to second order in the gradient expansion, requiring many more
coefficients than the celebrated shear viscosity to entropy ratio
$\eta/s$.  In particular, a treatment of collective flow requires not
only the shear viscosity $\eta$ but also the shear relaxation rate
$\taupi$, and baryon-number diffusion needs not just a diffusion
coefficient $D_q$ but also a diffusive relaxation time $\tauj$.

We would like to use experiments to constrain the properties of the
QGP such as $\eta/s$, but the necessity to include
higher-order coefficients could lead to a proliferation of fitting
parameters.  So one often assumes that the coefficients follow some
simple relations, such as $\taupi = K \eta/(\epsilon{+}\pressure)$, with
$(\epsilon{+}\pressure)$ the enthalpy density and $K$ a constant which we
draw from some microscopic theory of relativistic plasmas.  For
instance, Moore and York  showed that weakly-coupled massless QCD
treated to leading order (LO) in the gauge coupling
yields $5<K<6$ nearly independent of coupling strength \cite{York:2008rr}, while
Baier \textsl{et al} find that strongly-coupled $\mathcal{N}{=}4$ Super-Yang-Mills (SYM)
theory has $K \simeq 2.62$ \cite{Baier:2007ix}.

\begin{figure}[ht]
	\begin{center}
		\includegraphics[width=0.4\textwidth]{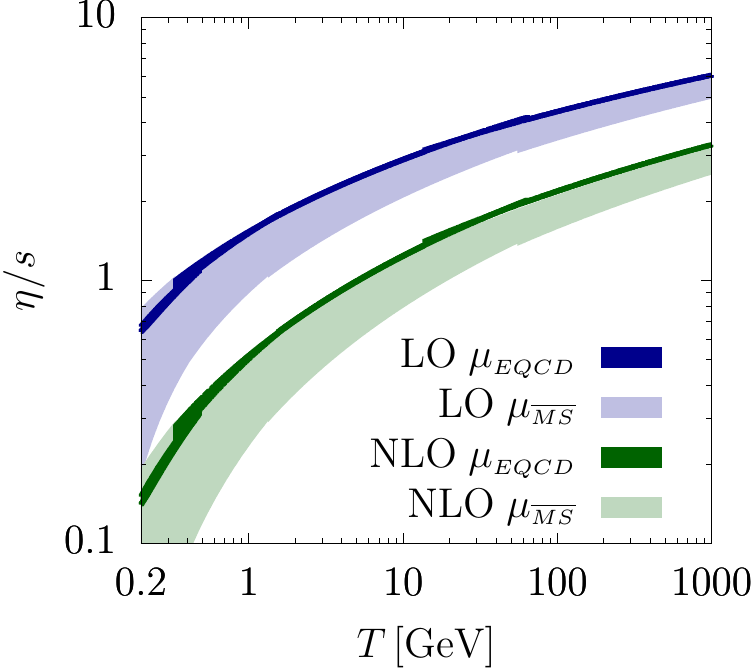}
      \end{center}
   \caption{$\eta/s$ of QCD
    as a function of temperature at LO 
           and NLO,  for the
          different choices of the running coupling
		   detailed in Fig.~\ref{fig_tau_T}
		  and in the text later on. Figure taken
		  from \cite{Ghiglieri:2018dib}.
        \label{fig:intro} }
\end{figure}

Recently we extended previous perturbative results for the shear
viscosity and baryon-number diffusion of hot QCD from leading \cite{Arnold:2003zc} to
next-to-leading order (NLO) \cite{Ghiglieri:2018dib}, see Fig.~\ref{fig:intro}.  
How does an NLO
treatment change $K$?  In this letter we will explore this issue.
Besides finding concrete results for
$K$ and $\tau_j/D_q$,
we will also show very general bounds
on these dimensionless ratios which follow as soon as we state that a
theory is well described by relativistic kinetic theory.  These bounds
are badly violated by strongly coupled theories with holographic duals, with the
interesting implication that these theories are \emph{very far} from
having quasiparticle descriptions.

\textbf{Definitions:}
Let us start by defining the objects of our investigation.
In the Landau-Lifshitz fluid rest frame 
the stress tensor has the form
\begin {equation}
    \langle T^{ij}(x) \rangle
    =
    \delta^{ij} \, \langle {\pressure} \rangle +\pi^{ij}\,,
	\label{eq:Tij}
\end{equation}
where the non-ideal dissipative part can be gradient-expanded. 
At first order
\begin{equation}
    \pi^{ij}_1=- \eta \left(\nabla_i \, u^j + \nabla_j \, u^i
		- {\textstyle \frac 23} \, \delta^{ij} \, \nabla_l \, u^l\right)
    - \zeta \, \delta^{ij} \, \nabla_l \, u^l \,.
\label{firstorder}
\end{equation}
We will concentrate on shear viscosity $\eta$ and not discuss bulk
viscosity $\zeta$ further.  At second order
the coefficients relevant for a conformal theory have been introduced
in \cite{Baier:2007ix}. Here we will only deal with second-order relaxation, 
whose coefficient $\taupi$ is defined as %
\footnote{Ref.~\cite{York:2008rr} showed that another second-order
  coefficient, $\lambda_2$ in the notation of \cite{Baier:2007ix},
  obeys $\lambda_2 = -2\eta\taupi$.  This still holds at NLO, so we
  determine this additional coefficient for free.}
\footnote{Ref.~\cite{Kovtun:2011np} shows that $\taupi$ should be
  understood as a Wilson coefficient; in the deep IR it receives
  formally divergent contributions from hydrodynamical fluctuations
  (long-time tails), which are small at all physically interesting
  time scales if $\eta/s > 0.2$.}
\begin{equation}
	\label{deftaupi}
	\taupi\partial_t \pi^{ij}=\pi^{ij}_1-\pi^{ij}.
\end{equation}

When there are additional
conserved global charges $Q_\alpha$
such as baryon or lepton number, the associated charge density
$n_\alpha \equiv j^0$ and current density $\j$
satisfy, at first order in the gradients, a diffusion equation,
\begin{equation}
    \langle {\j}_1 \rangle
    = -D_\alpha \> \naBla \langle n_\alpha \rangle = -\kmu \naBla \frac{\mu_\alpha}{T}  \,,
\label{eq:jia}
\end{equation}
where $D_\alpha$ is the diffusion coefficient.
Here
we have rewritten
the current with a gradient of the associated chemical potential $\mu_\alpha$. 
The associated transport coefficient $\kmu$ is 
related to $D_\alpha$ through
the susceptibility $\chi_\alpha$:
\begin{equation}
D_\alpha	= \frac{\kmu}{T\chi_\alpha}\,,\qquad \chi_\alpha\equiv\frac{\partial n_\alpha}{\partial \mu_\alpha}.
\label{sucept}
\end{equation}
If we were to write \Eq{firstorder} as a gradient of the charges $T^{0j}=(\epsilon+\pressure)u^j$, we would naturally
see that the associated relaxation coefficient is $\eta/(\epsilon+\pressure)$. 
Analogously to \Eq{deftaupi}, the second-order relaxation of $\j$ reads 
\cite{Betz:2008me,Hong:2010at,Denicol:2012cn}
\begin{equation}
	\label{deftauj}
\tauj\partial_t \j=\j_1-\j.
\end{equation}

In \cite{Arnold:2000dr,Arnold:2003zc} it was shown how the first-order
transport coefficients can be determined from a linearized kinetic theory.
In \cite{Arnold:2000dr} the collision operator defining the kinetic theory
of QCD was determined at leading logarithmic accuracy, in \cite{Arnold:2003zc}
at LO
and in \cite{Ghiglieri:2015ala,Ghiglieri:2018dib} at (almost) NLO.
The kinetic
theory expression for $\taupi$ was derived in \cite{York:2008rr}, leading
to its LO determination.

First we summarize
the main findings
of \cite{Arnold:2000dr,Arnold:2003zc,York:2008rr}.
We start from a generic kinetic theory of the form
\begin {equation}
     \left[
 	\frac{\partial}{\partial t}
 	+
 	\v_\p \cdot \frac{\partial}{\partial \x}
        \right]
f^a(\p,\x,t)
    =
    -C_a[f]\,,
\label {eq:Boltz}
\end {equation}
where $f^a(\p,\x,t)=dN^a/d^3\x d^3\p$ is the phase space distribution function for
the excitation (gluon, quark, antiquark) of index $a$.
If $u^i,\mu$ vary with space, then the local-equilibrium form of $f^a$
does as well %
\footnote{
   We use capital letters for four-vectors, bold lowercase ones
for three-vectors and italic lowercase for the modulus of the
latter. We work in the ``mostly plus'' metric, so that
$P^2=-p_0^2+p^2$. The upper sign is for bosons, and the lower sign is
for fermions.  The full collision operator is $C_a$; the collision
operator linearized in the departure from equilibrium is $\C_a$.},
$f_0^a=(\exp(-\beta u^\mu P_\mu - q^a_{\alpha} \beta\mu)\mp 1)^{-1}$.
The gradients on the left-hand side of \Eq{eq:Boltz}, which we treat as
perturbatively small, give rise to a source of departure from equilibrium,
$X^i=\nabla_i \mu_\alpha$ for flavor diffusion ($\ell=1$)
and $X^{ij}\propto\pi^{ij}_1/\eta$ for shear ($\ell=2$).  This
determines the linearized departure from equilibrium via a linearized
version of \Eq{eq:Boltz},
\begin{equation}
  \label{eq:Boltz1}
  \S^a(\p) = (\C f_1)^a(\p) \,,
\end{equation}
where $\S^a=\beta q^aX_{\ij} I_{\ij}(\p) f_0^a [1{\pm} f_0^a]$,
with $q^a = q^a_\alpha$ for number diffusion and $p$ for shear.
 $f_1$ is the linearized
departure from equilibrium,
$f^a(\p) = f_0^a(p) + f_1(\p) f_0[1{\pm}f_0]$ and
$I_{\ij} \propto p_i ..p_j$
(see \cite{Arnold:2000dr,Arnold:2003zc}).
At linear order $f_1^a \propto X_{\ij}$, allowing us to define
the scalar function $\chi(p)$ %
\footnote{This notation differs from that in
\cite{Arnold:2000dr,Arnold:2003zc,Ghiglieri:2018dib}.  $\S^a$ there is $-\S^a$ here,
and $\chi^a(p)$ there equals $-q^a\, \chi^a(p)$ here.}
\begin{equation}
  f_1^a(\p)
  \equiv \beta^2 X_\ij \, I_\ij(\hat\p) \:q^a\, \chi^a(p) \,.
\label{eq:f1}
\end{equation}
The linearized collision operator $\C$ is
worked out in detail for the case of weakly coupled QCD in
\cite{Arnold:2003zc} at LO and in \cite{Ghiglieri:2018dib} at NLO.

\textbf{General bounds:}
To determine $\eta$, $D_\alpha$, $\taupi$, and $\tauj$ we will need to
solve \Eq{eq:Boltz1} to linear order in $f_1$ but to subleading order
in gradients, which will depend in detail on the form of the collision
operator.  However we can already make some generic statements about
the solution, which will allow us to place bounds on certain
dimensionless ratios which hold automatically for all systems
described by relativistic kinetic theory, regardless of the details of
$\C$.  To see this, let us first define an inner product on the
Hilbert space of functions of momentum,
\begin{equation}
\label{eq:inner_product}
    \Big( g,h \Big) \equiv
    \beta^2 \sum_a \, \degen_a \! \int_\p \, (q^a)^2\,f_0^a(p)[1\pm f_0^a(p)]\, g^a(\p) \,
    h^a(\p) \,,
\end{equation}
with $\nu_a$ the degeneracy of species $a$ and $\int_\p\equiv\int\dpslash$.
Basic considerations such as stability ensure that the linearized
collision operator $\C$ is a linear, real, symmetric, positive semi-definite
operator under this inner product, and strictly positive in the channels
we consider.
In terms of this inner product, the first-order transport coefficients become
\cite{Arnold:2000dr,Arnold:2003zc}
\begin{equation}
	\label{defetakmu}
	\eta = \frac{1}{15}\Big( \chi,1 \Big),\qquad
	\kmu = \frac{T}{3}\Big( \chi,1 \Big).
\end{equation}
The enthalpy density and charge susceptibility can be easily obtained as
\begin{equation}
	\label{enthalpy}
	\epsilon+\pressure= \frac{T}{3}\Big( 1,1 \Big),\qquad
	\chi_\alpha= T\Big( 1,1 \Big).
\end{equation}

\begin{figure*}[ht]
	\begin{center}
		\includegraphics[width=\textwidth]{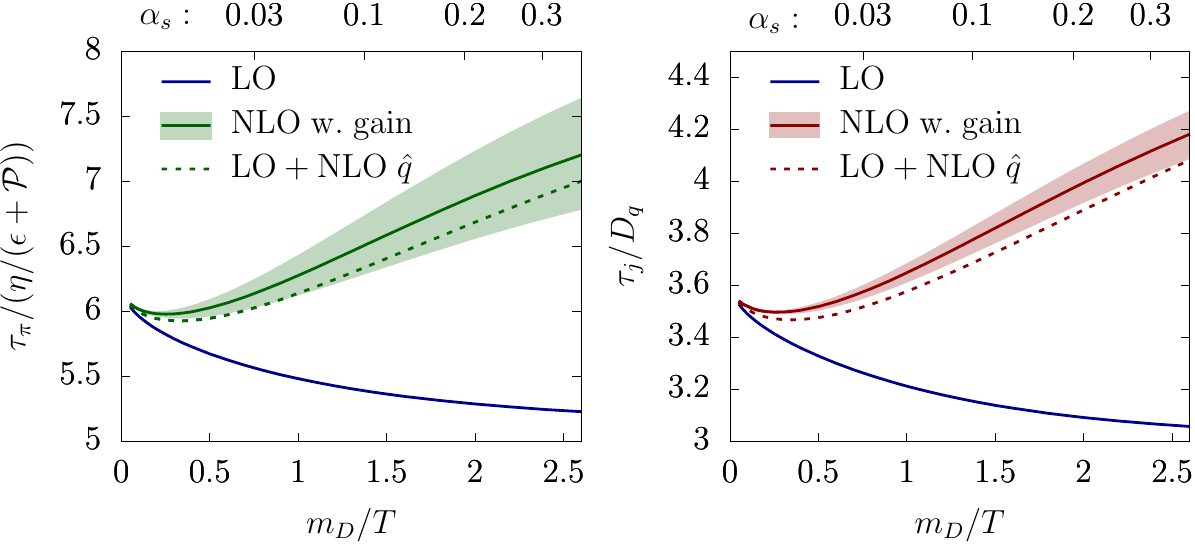}
	\end{center}
   \caption{
       The second- to first-order ratio of the relaxation coefficients for (a)  shear
       stress, $\taupi/(\eta/(\epsilon+\pressure))$, and (b) for  quark number diffusion, $\tauj/D_q$,
        as a function
   of $\md/T$ for  QCD with 3 light flavors. (The corresponding value
   of $\als$ is shown on the upper horizontal axis.) The LO result for $\taupi$ is
   from \cite{York:2008rr}, that for $\tauj$ is also new. The uncertainty from the unknown gain
   terms is shown by the bands; it is estimated as specified in \cite{Ghiglieri:2018dib}
   by the LO value for the gain terms, times $\md/T$, times a constant in 
   the interval $[-2,2]$.
   The dashed lines represent an estimate in which we include only the
   NLO $\qhat$ to the LO collision operator.
      \label{fig_tau_md} 
   }
\end{figure*}
$\taupi,\tauj$
require inserting $f_1$ into
the left-hand side of \Eq{eq:Boltz} and using the time derivative to
find $f_2$ at one space-derivative, one time-derivative order.
As shown in \cite{York:2008rr}, the properties of the inner product
and of $\C$ then turn the evaluation into
the inner product of the first-order departure from equilibrium $\chi$
with itself:
\begin{equation}
	\label{kintaupi}
 	\eta \taupi = \frac{\beta}{15}\Big( \chi,\chi \Big).
\end{equation}
The same analysis can be applied to $\tauj$ and we find 
\begin{equation}
	\label{kintauj}
 	\kmu \tauj = \frac{1}{3}\Big( \chi,\chi \Big).
\end{equation}
It is then insightful to consider these 
dimensionless ratios,
\begin{equation}
\label{secondorder}
\frac{\taupi}{\eta/(\epsilon+\pressure)}
=5\frac{\Big( \chi,\chi \Big)\Big( 1,1 \Big)}{\Big( \chi,1 \Big)^2},\qquad
\frac{\tauj}{D_\alpha}=3\frac{\Big( \chi,\chi \Big)\Big( 1,1 \Big)}{\Big( \chi,1 \Big)^2},
\end{equation}
which also have the same number of powers of the collision operator
($\chi \propto \C^{-1}$) in the numerator as in the denominator.
The triangle inequality implies
\begin{equation}
\label{bounds}
\frac{\taupi}{\eta/(\epsilon+\pressure)}
\ge 5\,,\qquad
\frac{\tauj}{D_\alpha}\ge 3\,.
\end{equation}
These results apply to any kinetic theory description of these transport coefficients,
as long as the enthalpy density or the charge susceptibility are also 
consistently computed within the kinetic theory.
We remark that
the $\ell=1,2$ departures from equilibrium contributing to these transport coefficients
do not by construction contribute to the ($\ell=0$) thermodynamical 
functions $\epsilon+\pressure$ or $\chi_\alpha$.

In contrast, strong-coupling results from the AdS/CFT correspondence
in $\mathcal{N}{=}4$ SYM theory
give for $\taupi$ \cite{Baier:2007ix}
and for the relaxation of a $U(1)$ current in SYM  \cite{Bu:2015ame}
\begin{equation}
	\label{strongcoupling}
	\frac{\taupi}{\eta/(\epsilon+\pressure)}\bigg\vert_\mathrm{AdS}=4-2\ln(2)\,,\qquad
	\frac{\tauj}{D_{U(1)}}\bigg\vert_\mathrm{AdS}=\frac\pi 2\,.
\end{equation}
In both cases, these strong-coupling results are approximately
half the minimum value attainable in kinetic theory. Finite-coupling
corrections \cite{Buchel:2004di,Benincasa:2005qc,Buchel:2008ac,Buchel:2008wy,Buchel:2008bz} to the first ratio show a modest increase.
We also note that our kinetic theory bounds in \Eq{bounds} can be shown to become, in $d$
spatial dimensions, $d+2$ and $d$ respectively. It would be interesting 
to derive larger-dimension holographic results in comparison.

\begin{figure*}[ht]
	\begin{center}
	\includegraphics[width=\textwidth]{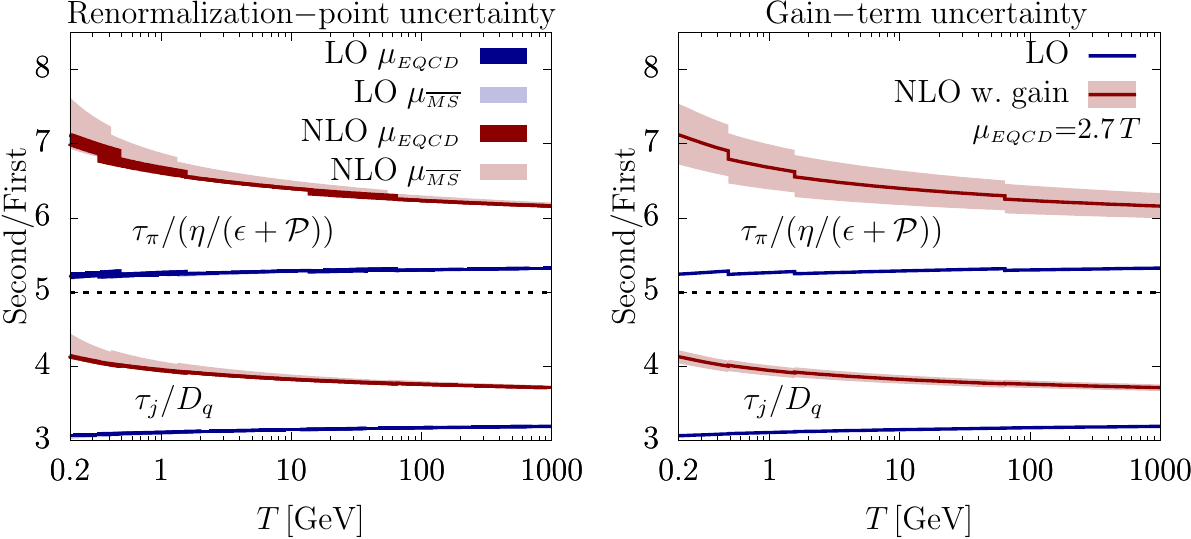}
	\end{center}
   \caption{The second- to first-order ratio of the relaxation coefficients for  shear
       stress, $\taupi/(\eta/(\epsilon+\pressure))$ (values above 5), and for  quark number diffusion, $\tauj/D_q$
	   (values below 5),  as a function of $T$. 
	   On the left, we plot different choices of the running coupling:
	 the solid bands fix the coupling using the two-loop EQCD value with
 $\mu_{EQCD}=(2.7\leftrightarrow 4\pi) T$, while the shaded bands
     use the standard $\overline{\mathrm{MS}}$
       two-loop coupling with $\mu_{\overline{\mathrm{MS}}}=(\pi \leftrightarrow 4\pi)T$. 
	   On the right we plot
	   instead in the shaded red bands the estimated uncertainty due to
           the gain terms.
   	  All curves in this plot are obtained using the effective
     EQCD coupling with $\mu_{EQCD}=2.7\,T$. 
   \label{fig_tau_T} }
\end{figure*}

\textbf{Second-order relaxation at (almost) NLO:}
We now provide results for the second-order relaxation of the shear stress
tensor and of the light quark current $j_q$ in QCD. In \cite{Ghiglieri:2018dib}
we have introduced in great detail a linearized collision operator to 
``(almost) NLO''.  (Corrections which lie beyond the kinetic-theory
picture arise at still higher order.)
$\twotwo$ elastic scatterings and effective 
$\onetwo$ inelastic scatterings contribute to the LO collision operator, the
former taking the lion's share.
At NLO we found all new scattering
processes, and corrections to the LO processes, which
are suppressed by a single power of the QCD coupling $g$. 
As we showed in detail, there are only a
few such $\OO(g)$ subleading effects.  First, the rate of soft
$\twotwo$ scattering is modified; this can be described as an
additional momentum-diffusion coefficient $\delta\qhat$.  This
modification, and an $\OO(g)$ correction to the in-medium 
dispersion, also provide an $\OO(g)$ shift in the $\onetwo$ 
rate.  Next, this $\onetwo$ splitting rate must be corrected wherever
one participant becomes ``soft'' ($p \sim gT$) or when the opening
angle becomes less collinear.  And finally, subtractions are needed
because of the way the numerical
implementation of the LO scattering kernel \cite{Arnold:2003zc}
already resums a small amount of the NLO effects.
We were able to give a relatively simple determination of these effects
by the use of light-cone techniques fostered by \cite{CaronHuot:2008ni}. 
 Unfortunately, these methods
typically keep track of the incoming and outgoing momentum of a
particle, but lose track of the momentum which it transfers to the
other participants.  This momentum transfer also affects the departure
from equilibrium of the other particle or particles which receive the
momentum, generating, in the effective Fokker-Planck approach applicable
for these soft scatterings, a \emph{gain term}. This is
an effect which we  failed to account for at NLO, hence the ``almost'' NLO.  
However we estimated
that  this missing part is most likely small. Finally, we found
out that  $\eta/s$ and $D_q$ at NLO become smaller than their
LO counterparts by a factor of 4 at the couplings of relevance
for heavy ion collisions, see Fig.~\ref{fig:intro}. 
The large $\delta\qhat$ contribution is
by far the main contribution responsible for this behavior.

We now use this (almost) NLO collision operator to determine
$\taupi$ and $\tauj$ using Eqs.~\eqref{eq:Boltz1}
and \eqref{secondorder}. We solve Eq.~\eqref{eq:Boltz1} 
with the same variational method as in 
\cite{Ghiglieri:2018dib}, which also details the NLO
operator $\delta\C$.
In Fig.~\ref{fig_tau_md} we plot our results for the
second-order coefficients $\taupi$ and $\tauj$, normalized as in Eq.~\eqref{secondorder},
as functions of the Debye mass $\md\sim gT$ over the temperature. 
The LO results for $\taupi$ were originally obtained in
\cite{York:2008rr}. Those for $\tauj$ are new and
consistent with the leading-log estimate in \cite{Hong:2010at}.
The plot shows that both LO results in solid blue decrease with increasing coupling,
approaching the minimum values (\Eq{bounds}), while the NLO results in solid green
and red respectively start to differ significantly from the LO at
$\md\simg 0.5 T$, where they start growing, getting in the ballpark of
$3/2$ of the minima when $\als\sim 0.3$.  The dashed green/red
curves are the results obtained by adding only $\delta\qhat$ to the LO
collision operator, showing that also in this case it dominates NLO
corrections. The bands are obtained by varying the estimate for the
unknown gain terms within a range reasonably encompassing their
probable size (and sign), as described in \cite{Ghiglieri:2018dib}.
Intuitively, the LO results approach the bound at increasing coupling
because the log-enhanced $\twotwo$ processes, which force $\chi(p)\propto p$,
become less effective at larger couplings, while the other processes
drive $\chi(p)$ to a constant, saturating \Eq{secondorder}. At NLO
the large $\delta\qhat$ drives $\chi(p)$ towards $p^2$, which is further
from the bound.

Fig.~\ref{fig_tau_T} presents the more phenomenologically relevant
dependence of these second-order coefficients on the temperature.
Since only an NNLO treatment would directly include running-coupling
effects, this requires that we pick a prescription for relating the
running coupling to the temperature.  We do so by either using the $\MS$
coupling in the range $\pi T < \mu_{\MS} < 4\pi T$ (leading to the
larger, light-shaded bands in the left plot) or via the
effective Electrostatic QCD (EQCD) coupling with $2.7 \, T < \mu_{EQCD} < 4\pi T$
as in Ref.~\cite{Laine:2005ai} (narrow, dark-shaded bands in the left
plot).  The discontinuities in the plot occur where we change
prescriptions for the number of light fermion species.
The right plot in the figure indicates the errors due to the
uncertainties from our ignorance of the gain terms which we
discussed above.

\textbf{Conclusions:}
Viscous hydrodynamical studies of heavy ion collisions require
second-order hydrodynamical coefficients $\taupi$, $\tauj$ which can be
understood as relaxation times towards the first-order behavior.
While the hydrodynamic coefficients such as $\eta/s$ and $\taupi T$ vary
by orders of magnitude as a function of temperature and differ
substantially between LO and NLO calculations (see Fig.~\ref{fig:intro}), 
we have shown that simple dimensionless ratios, \Eq{secondorder}, are remarkably
robust, varying  at most by $40\%$ as a function of
coupling/temperature and between LO and NLO determinations.
Furthermore and more remarkably, we have shown that in \emph{any}
theory which can be described by kinetic theory of ultra-relativistic
particles, these dimensionless ratios obey inequalities,
shown in \Eq{bounds}.  These inequalities hold regardless of the
details of the collision operators, and they give the hydrodynamics
practitioner a simple prescription for how to estimate the relation
between first-order and second-order transport coefficients.

It is also remarkable that the bounds we find fail by a full factor of
2 when we compare them to the results within strongly coupled theories
with holographic duals.  We conclude that such strongly coupled
theories are very far from  having a kinetic description.  This
provides a useful counterpoint to the frequent unspoken assumption
that the QGP should have a kinetic description.

\begin{acknowledgements}
\textbf{Acknowledgments:}
JG would like to thank Aleksi Kurkela
for useful conversations.
GM would like to acknowledge support by the Deutsche
Forschungsgemeinschaft (DFG) through the grant CRC-TR 211
``Strong-interaction matter under extreme conditions.''
DT would like to acknowledge support by the U.S. Department of Energy
through the grant DE-FG02-88ER40388.
%
%
\end{acknowledgements}

\bibliography{eloss}

\end{document}